# SUBMACROPULSE ELECTRON-BEAM DYNAMICS CORRELATED WITH HIGHER-ORDER MODES IN A TESLA-TYPE CRYOMODULE

A. H. Lumpkin[#], R. Thurman-Keup, D. Edstrom, P. Prieto, J. Ruan
Fermi National Accelerator Laboratory, Batavia, IL 60510 USA

B. Jacobson, J. Sikora, J. Diaz-Cruz[1], A. Edelen, F. Zhou,
SLAC National Accelerator Laboratory, Menlo Park, CA 94720 USA
[1]also at University of New Mexico, Albuquerque, NM 87131 USA

## ABSTRACT

Experiments were performed at the Fermilab Accelerator Science and Technology (FAST) facility to elucidate the effects of long-range wakefields (LRWs) in TESLA-type superconducting rf cavities. In particular, we investigated the higher-order modes (HOMs) generated in the eight cavities of a cryomodule (CM) due to off-axis steering with correctors located ~4 m upstream of the CM. We have observed correlated submacropulse centroid slews of a few-hundred microns and centroid oscillations at ~240 kHz in the rf BPM data after the CM. The entrance energy into the CM was 25 MeV, and the exit energy was 100 MeV with 125 pC/b and 400 pC/b in 50-bunch pulse trains. These experimental results were evaluated for machine learning training aspects which will be used to inform the commissioning plan for the Linac Coherent Light Source-II injector CM.

Index: 41.60

[#]lumpkin@fnal.gov

# I. INTRODUCTION

The preservation of the low emittance of electron beams during transport in the accelerating structures is an ongoing challenge for large facilities. In the cases of the TESLA-type superconducting rf cavities currently used in the European X-ray Free-electron Laser (EuXFEL) [1] and the under-construction Linac Coherent Light Source upgrade (LCLS-II) [2], off-axis beam transport may result in emittance dilution due to transverse long-range wakefields (LRWs) and short-range wakefields (SRW) [3-5]. To investigate such effects, experiments were performed at the Fermilab Accelerator Science and Technology (FAST) facility with its unique configuration of two TESLA-type cavities after the photocathode rf gun followed by an 8-cavity cryomodule [6]. Previously we reported beam dynamics studies in regard to the single TESLA-type cavities for LRWs [4] and SRWs [5], but here we extend the LRW techniques to a full 8-cavity cryomodule for the first time. In addition, we also obtained detailed spectrally-resolved HOM data with a high-resolution oscilloscope using the FFT of time-domain data. This enabled us to track the 18 individual dipolar modes in the first two passbands *including phase and mode polarization effects* [7]. Further, as a developmental activity, we used the steering-dependent HOM and rf BPM data for machine learning (ML) training in regard to minimization of HOMs and reduction of emittance dilution effects in a CM [8]. This result will be relevant to LCLS-II injector commissioning where < 1-MeV beam will be injected into the first cryomodule with the gradients of the first 3 cells being 8, 0, and 0 MV/m and the last 5 cells at 16 MV/m [9]. Wakefield kicks to the beam go inversely as beam energy so steering on axis into these first cavities will be crucial to avoid emittance dilution [4,5].

We generated beam trajectory changes with the corrector magnet set located 4 m upstream of the cryomodule. We observed correlations of the cavity higher-order modes' (HOMs) signal levels and submacropulse centroid slews and/or centroid oscillations in the 11 BPM locations downstream of the cryomodule. At 125 pC/bunch, 50 bunches per macropulse, 25-MeV entrance energy, and 100-MeV exit energy, we observed for the first time submacropulse position slews of up to 500 microns at locations within 1.2 m after the CM and a centroid oscillation at a difference frequency of 240 kHz further downstream. Both are emittance-dilution effects which we mitigated with selective upstream beam steering. The experiments were facilitated by the implementation of two prototype HOM detector chassis designed for the LCLS-II injector cryomodule. These chassis were based on the basic Fermi National Accelerator Laboratory (FNAL) design with a dipolar mode passband and zero-bias Schottky detectors, and they had additional selectable features for up to two wideband amplifiers in series and attenuators in each channel [10]. The combined 8 channels then enabled signal acquisition from one coupler of all eight cavities at the same time to assess the beam trajectory relative to the cavity-mode centers. In addition, we fielded a hybrid HOM detector box which selectively filtered the raw signal at two additional frequency bands centered at 2.5 and 3.25 GHz. These hybrid boxes also had raw, filtered outputs which were used with the high-speed oscilloscope to assess steering effects and resolve frequency splitting of the two mode polarizations in the two different cavity structures from two vendors.

## II. EXPERIMENTAL ASPECTS
### A. The FAST Electron Linac

The Integrable Optics Test Accelerator (IOTA) electron injector at the FAST facility (Fig. 1) begins with an L-band rf photoinjector gun built around a $Cs_2Te$ photocathode (PC). When the UV component of the drive laser, described elsewhere [11] is incident on the PC, the resulting electron bunch train with 3-MHz micropulse repetition rate exits the gun at <5 MeV. Following a short transport section with a pair of trim dipole magnet sets, the beam passes through two superconducting rf (SCRF) capture cavities denoted CC1 and CC2, and then a transport section to the low-energy electron spectrometer. In this case this dipole power is off so 25-MeV beam is transported to and through the CM with an exit energy of 100 MeV. The beam parameters are shown in Table I. The micropulse numbers were either 1 or 50. Diagnostics used in these studies included the rf BPMs in the beamline (plus the cold BPM at the downstream end of the CM) and the HOM couplers/detectors at the upstream (US) and downstream (DS) ends of each of the 8 SCRF cavities.

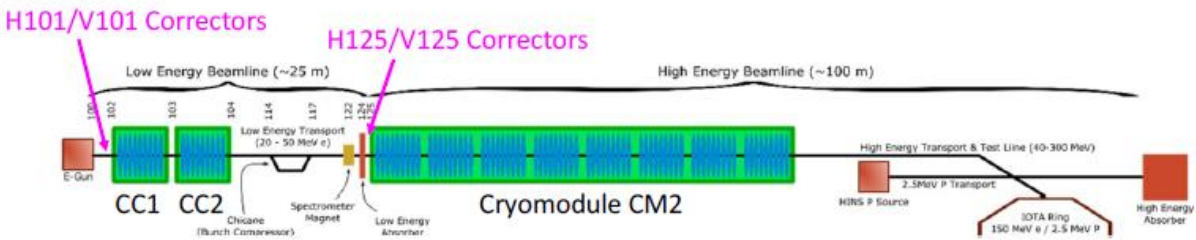

Figure 1: FAST Linac showing the location of the CC1 and CC2 single cavities, and the cryomodule CM2 with 8 cavities. The corrector locations are also shown just upstream of the cavity locations. This schematic is not to scale.

Table I: Summary of the electron beam parameters used in the CM studies.

| Beam Parameter | Units | Value |
|---|---|---|
| Beam Energy | MeV | 25-100 |
| Charge | pC | 125-500 |
| Bunch Length (rms) | ps | 10-15 |
| CC1 rf Gradient | MV/m | 21 |
| CC2 rf Gradient | MV/m | 0 |
| CM:Cn Gradient | MV/m | ~9.5 |

### B. The HOM Detectors

There are 16 HOM outcouplers within the CM. The HOM signals were processed by the HOM detector circuits with the Schottky diode output provided online though ACNET, the Fermilab accelerator controls network [6]. The HOM signals are transported from the cavities to the electronics via 1/2" Heliax cables. Once in the rack, they transition to BNC-terminated RG-58 cables for convenience before entering one of several analog filter boxes. Since there were different filter boxes being tested, a patch panel was used to quickly switch between cavities and boxes. Unlike the downconverter electronics design with a narrow band filter on the $TE_{111}$ mode used by others [12,13], the HOM detectors' bandpass filters were optimized for two dipole passbands from 1.6 to 1.9 GHz, and the 1.3-GHz fundamental was reduced with a notch filter.

With the implementation of the two SLAC prototype HOM chassis with 8 channels in total (after commissioning on CC1 [9]), we could cover all 8 US or 8 DS couplers at one time as schematically shown in Fig. 2. In addition, the hybrid FNAL HOM chassis provided three filtered outputs of two channels on the unused C1 and C8 outputs as indicated. Two 12-channel, 10-bit digitizers provided by FNAL established the link to the ACNET system. The rf BPMs' electronics were configured for bunch-by-bunch capability with optimized system attenuation for each charge [14].

The schematic of the hybrid HOM filter box used in this set of measurements is shown in Fig. 3. It contains two channels, with each channel having three bandpass filters each with a passband of ~0.3 GHz centered at 1.75 GHz, 2.5 GHz, and 3.25 GHz, and an optional 20 dB amplifier controlled by a switch on the front panel. Three of the outputs (one of each frequency band) pass through a Schottky diode to rectify the signal before sending it to the digitizers in the control system. The other three outputs were connected to an 8-GHz analog bandwidth, 20-GS/s, Rohde & Schwarz RTP084 oscilloscope which recorded the HOM signals for further processing offline. The initial DC block in the circuit was to prevent any charge buildup from getting to the 1.3-GHz notch filter. The proximity to the beam harmonics is important when considering build-up of the HOMs along the bunch train.

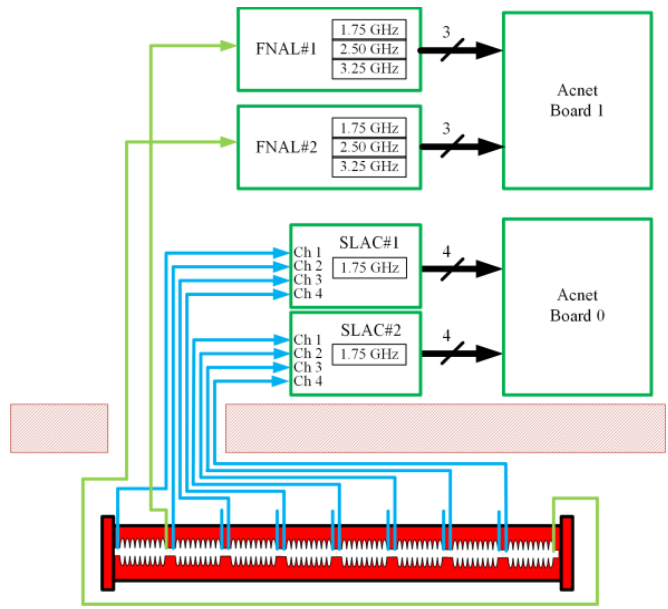

Fig. 2. Schematic of 8 channels from the US couplers of 8 cavities plus FNAL channels on C1 and C8 DS channels going into the digitizer boards [9].

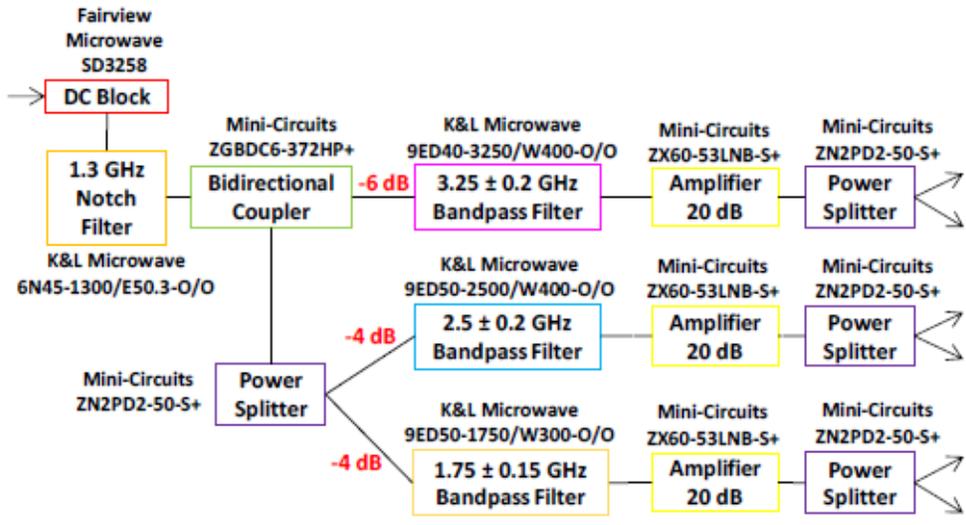

Figure 3: Schematic of the hybrid HOM filter circuit. The red dB numbers indicate the relative amplitudes of the three channels after the combined bidirectional coupler and power splitter [7].

55

## III. EXPERIMENTAL RESULTS

### A. Initial CM2 HOM waveform data: 125 pC/b

An assessment of beam offset in the CM2 cavities was provided by the HOM data, both upstream (US) and downstream (DS) signals. Examples of initial digitized waveforms displayed in real time during the experiment are shown in Fig. 2. The CM2 HOM signals were surprisingly weaker than those of the single cavities, CC1 and CC2 at this charge. When no amplifiers were used, only the C3 and C4 US HOM waveforms, green and yellow respectively, are noticeably above the baseline in Fig. 2a. The use of a single wideband amplifier on each channel made all 8 signals measurable in Fig. 2b. Steering with corrector V125 located just 4 m before CM2 (see Fig.1) resulted in the correlated trajectory and HOM signal changes. The steering angle was ~ 2 mrad/A into CM2. Our reference currents were H125=1.5 A and V125=4.3 A. Peak-value data with the H125 corrector current scans are shown in an initial report [10] and in Fig. 5.

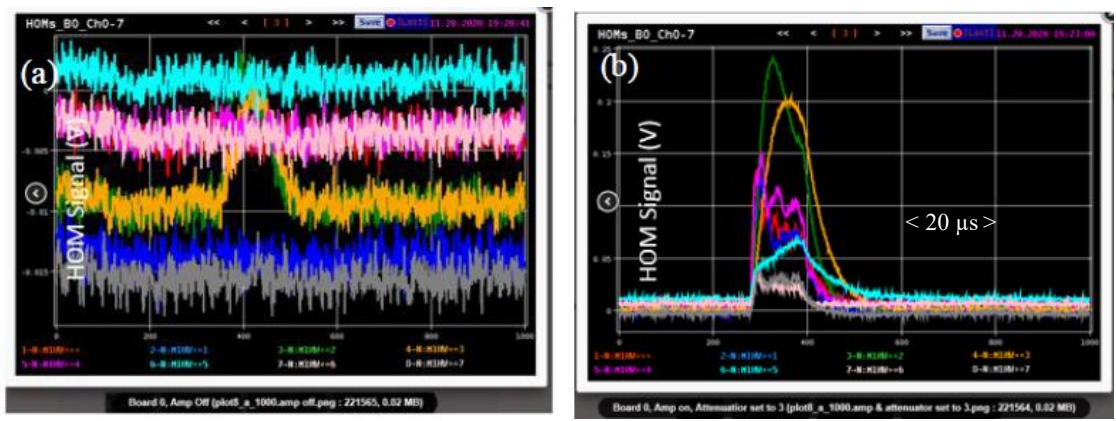

Figure 4: (a) Initial US HOM waveforms for Cavities 1-8 with colors red, blue, green, yellow, purple, light blue, pink, and grey, respectively. Only C3 and C4 seem to have usable signals with no amplifier. (b) The US HOM signals with one wideband amplifier switched on. All 8 Schottky detector signals are now measurable on the same shot. These are the dipolar modes using 50 b over a 16.6 µs temporal extent.

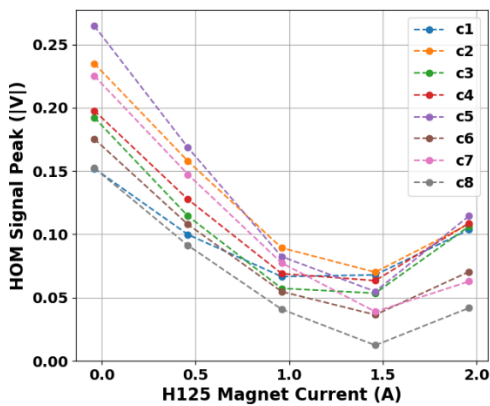

Figure 5: Initial plots of the CM2 HOM US peak signals vs H125 corrector current values. One amplifier is used in each channel with charge at 125 pC/b and 50 b.

## B. Bunch by Bunch rf BPM Data: 125 pC/b

We also evaluated the potential long-range wakefield or HOM effects on the beam centroid by using the rf BPM bunch-by-bunch data after CM2. These data were obtained in an off-normal condition with CC2 tuned 15 kHz off-resonance and powered off. This is discussed elsewhere, but we did identify a residual submacropulse beam-centroid oscillation at ~220 kHz due to CC2 [14].

The beam energy was 25 MeV entering CM2 and 100 MeV exiting. An example of the centroid motion within the 50-micropulse train in a macropulse is shown in Fig. 6 with both noise-reduction and bunch-by-bunch capabilities implemented. The first three vertical BPMs after CM2:C8 show an increasing centroid slew correlated with V125 settings which grows to 500 μm in B441. B418 is the cold BPM located just after C8 and still within the CM. Further downstream we show samples from B480 where a vertical centroid oscillation of ~240 kHz is seen in Fig. 7 for the two larger -1 A V125 corrector values. In this preliminary set, we did not have an offset minimum established at 0.0 A from the reference although the oscillation amplitude is noticeably larger at -1 A (-2 mrad). We attribute this to the combined difference frequencies between the dipolar HOMs and the beam harmonics in the 8 cavities such as found in subsection E.1 and reference [4].

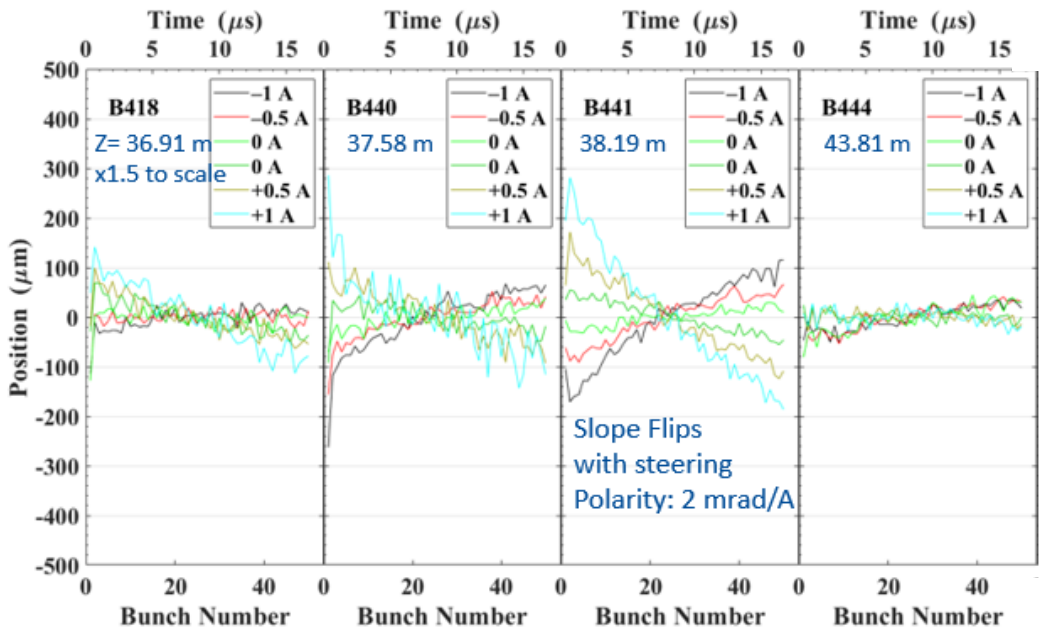

Figure 6: Submacropulse vertical centroid effects in the first four BPM locations after CM2:C8 correlated with V125 settings. This includes the cold BPM denoted B418. The z-locations are shown in each panel with the maximum centroid slewing seen in B441. The V125 corrector steering angle change is 2 mrad/A.

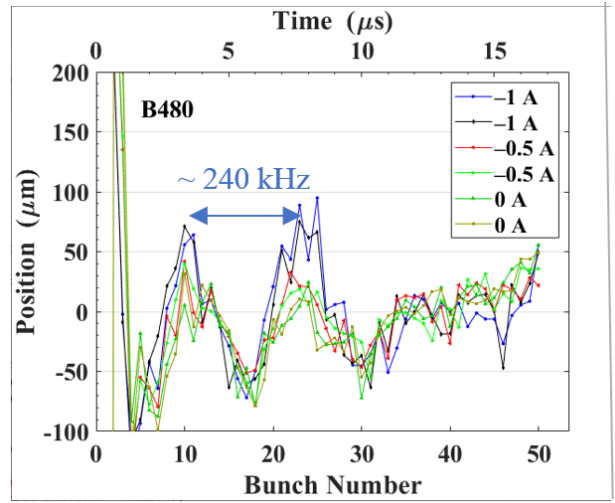

Figure 7: Examples of the variation with V125 corrector current of the beam vertical centroids bunch by bunch for 50 micropulses at B480. The ~240 kHz centroid oscillation is indicated.

### C. Integrated HOM Schottky Data: 400 pC/b

For the V125 corrector scan we show the summary results in Fig. 8 for all 16 HOM digitized, frequency-integrated dipolar channels centered at 1.75 GHz. We performed the V125 current scan by setting the corrector current step and acquiring data for the 8 US channels at each charge and then for the 8 DS channels at each charge. The relative minima are seen for all 16 channels at our reference setting with 400 pC/b, but we observe that the sensitivity to steering is greater for the early cavities than the DS ones. Cavity 3 (green curve) and C8 (black curve) seem to have elevated minima in the US signals indicating a larger beam offset to the cavity dipole modes' centers. From the amplitude minimum seen in the spectral data for CM2:C1 dipole mode 7 as well, we deduce this effect in Fig. 8a is probably due to the C3 and C8 misalignments relative to the adjacent cavities. There may be a vertical cavity tilt along their z axis.

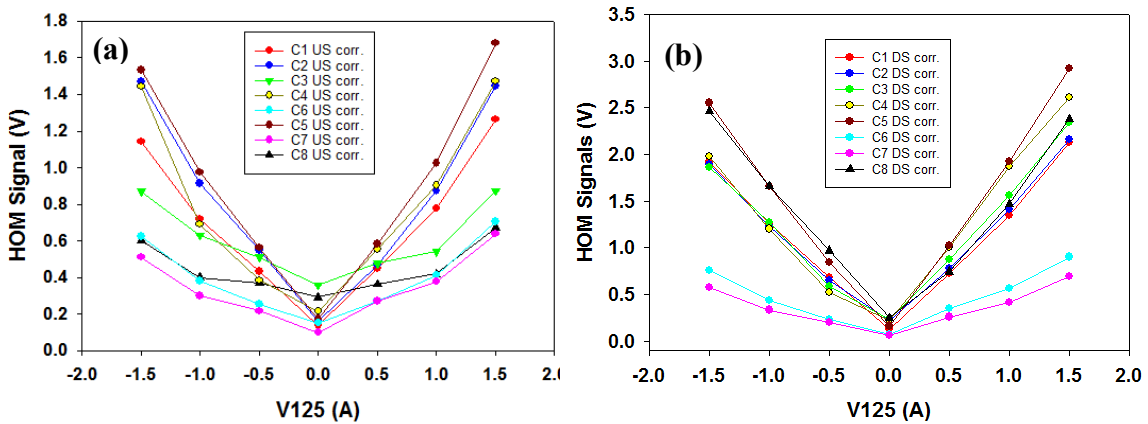

Figure 8: Measurements of the CM2 a) US and b) DS HOM peak signals from the Schottky data. We have identified a reference corrector current setting for our 0.0 A with an actual V125 current value of 4.3 A after scanning for a relative minimum in CM2 US HOM channels. The H125 value was fixed at a reference.

By correlating the B130 BPM readings with V125 corrector steering and the CM2:C1 US HOMs, one can generate a beam-offset monitor (BOM) plot at the entrance of CM2 with 400 pC/b and 50 b as in Fig. 9. Note the horizontal axis is in mm for this plot. These data were obtained for the first three cavities with 1 amplifier enabled. Again a comparison of the beam offset values for the first three cavities indicates C3 has a noticeable beam offset of ~ 2 mm at its US HOM coupler location. This is the same cavity type as C1 for vendor source while C2 is from the other vendor.

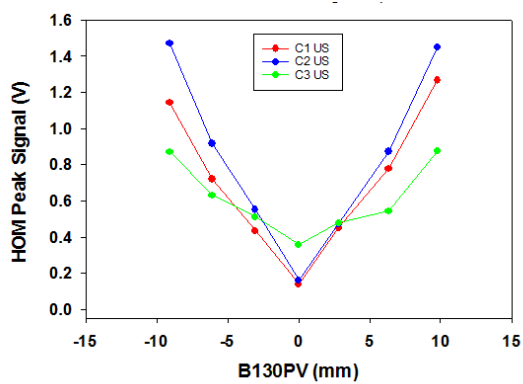

Figure 9: Beam-offset-monitor plot with 400pC/b at entrance to CM2 and the first 3 cavities.

### D. Integrated HOM Schottky Data: Single Bunch 10 pC

An important test for the SLAC prototype chassis is its ability to detect offsets with a single bunch of 10 pC. Figure 10 shows the signals obtained from all sixteen HOM probes with a single 10 pC bunch without changing the accelerator magnet settings. These data were taken after minimizing the upstream signals by hand adjustment of steering magnet settings. Both cascaded amplifiers in each channel are enabled with no added attenuation.

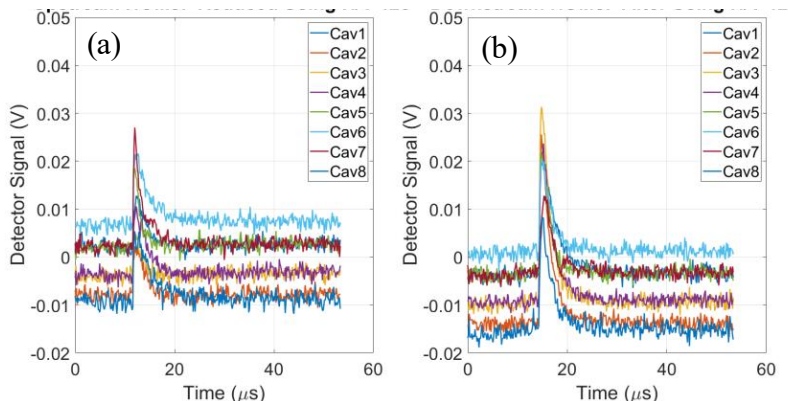

Figure 10: Output of the SLAC chassis with a single bunch of 10 pC and the beam approximately on axis for a) US HOMs and b) DS HOMs (see text). Both cascaded amplifiers are enabled in all channels [9].

## E. HOM Spectral Data: 400 pC/b

### E.1  1.75-GHz Passband Data

In a new initiative for us, we operated with a single bunch to explore the detailed spectral information obtainable with a Rohde & Schwarz 20 GS/s oscilloscope [7]. The filtered HOM signals near 1.75 GHz were directly recorded for 15-20 µs after the beam trigger for the 18 dipolar modes in the passband as shown in Figs. 11a and 11b, respectively. Figure 11c shows the spectrum near the 1.73-GHz HOM in 4 cavities, the measured mode values using a network analyzer connected between the two couplers of the cavity (triangles), and the 3-MHz beam harmonics (red circles). Figure 12 shows an example of the FFT of the temporal digital data near 1.72 GHz for CM2:C1 US dipole mode 7. In this case, we resolve the horizontal and vertical polarization components with the horizontal component at 1.7266 GHz clearly changing during the H125 scan from -1 A (-2 mrad) to +1 A (+2 mrad) and the vertical component at 1.7241 GHz changing during the V125 scan. This indicates the mode polarization axes are aligned with the horizontal and vertical spatial axes. The frequency splitting is more than 2 MHz for this Advanced Engineering Systems (AES)-built cavity. The cavity is not axisymmetric evidently. As another example we show the CM2:C8 mode 7 near 1.730 GHz in Fig. 13. In this Research Instruments (RI)-built cavity, the frequency splitting is ~1 MHz, and only the vertical component at the lower frequency changes amplitude with the V125 scan values in Fig. 13b.

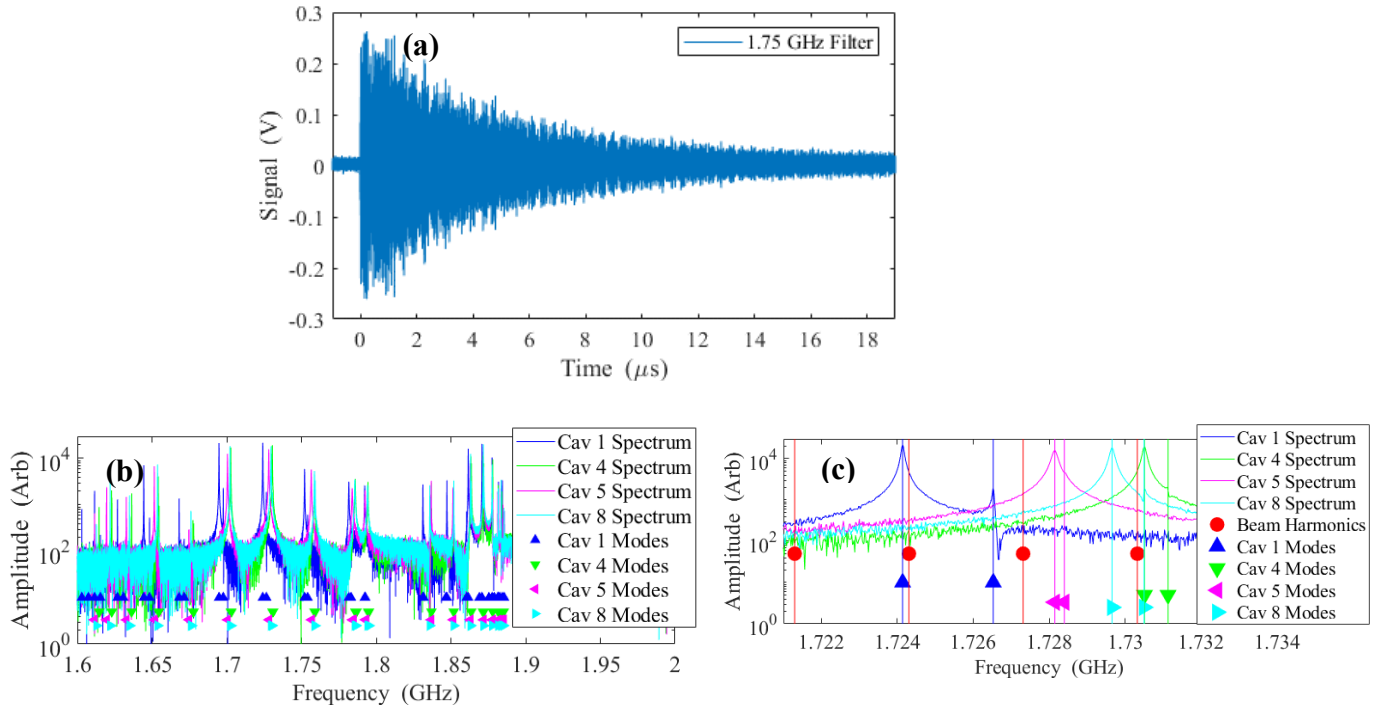

Figure 11: a) HOM signals after 1.75-GHz passband filter from cavity 1 of CM2. b) Spectra of 4 cavities in CM2. c) Dipolar Mode-7 frequencies can vary by up to several MHz in the 4 sampled cavities of the CM, and the C1(blue) and C4 (green) cavities have near resonances with the beam harmonics (red circles).

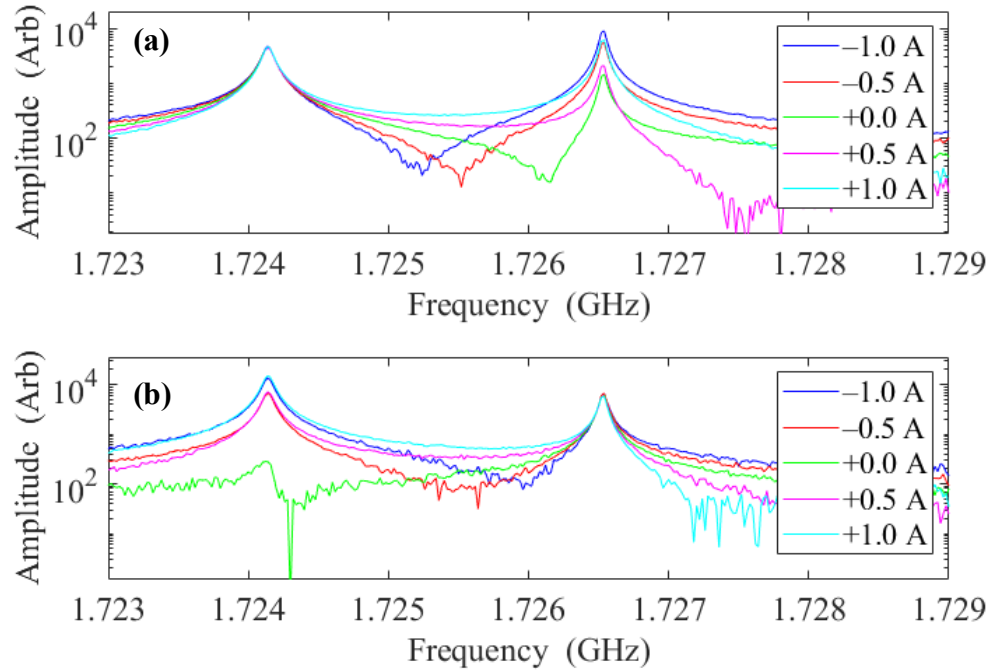

Figure 12: Horizontal (a) and vertical (b) steering results showing polarization-dependent HOM variations and ~2-MHz frequency spitting of the two polarizations of dipolar mode 7 in CM2:C1 (AES).

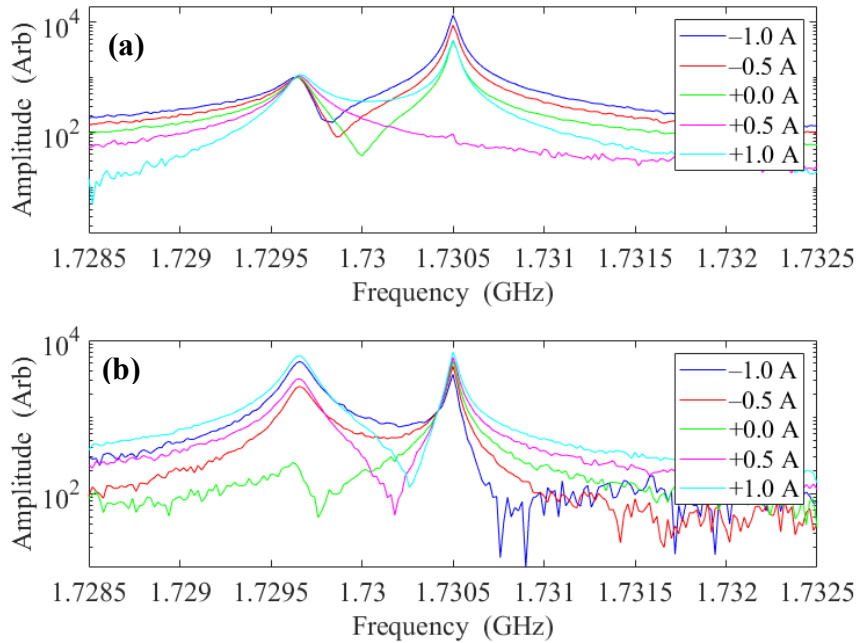

Figure 13: Horizontal (a) and vertical (b) steering results showing polarization-dependent HOM variations and ~1-MHz frequency spitting of the two polarizations of dipolar mode 7 in CM2:C8 (RI). Only the vertical polarization component at 1.7297 GHz is responsive to the V125 scan.

To correlate the HOMs with beam steering direction requires the phase of the HOM as well as the magnitude. Extracting the phase requires a stable trigger. For these measurements, we used a trigger for the oscilloscope that was stable enough to allow a time correction to be applied later. Figure 14a shows the start of the HOM signals which should be dominated by the arrival time of the bunch at the HOM pickup. One can see variation in the arrival time which needs to be corrected to get the correct phase of the HOMs. Figure 14b shows the signals after adjusting for the jitter using the peak located at $t = 0$ ns. In Figure 15 we see the Mode 7 amplitude in CM2:C1 for the V125 corrector scan. In this case we applied the phase information so we extract the sign of the vertical steering. This confirms our assessment from the HOM Schottky data for minimized HOM dipolar signals shown in a previous subsection, III-C.

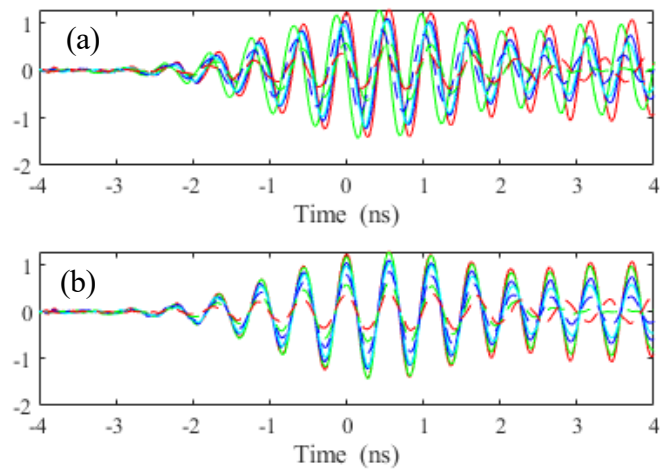

Figure 14: a) HOM signals from the oscilloscope. b) After aligning time signals using the peak at 0 ns.

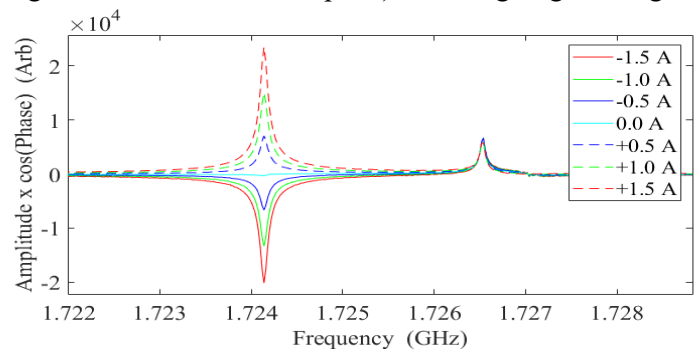

Figure 15: 'Signed' spectra of ~1.72 GHz mode showing the V125 vertical corrector steering effects on the dipolar mode 7 amplitude in C1. The reference 0.0-A setting is confirmed with the low mode amplitude observed in this and for other dipolar modes.

### E.2    2.5-GHz Passband Data

The objective of acquiring data in this regime was to determine if the dipolar Mode 30 at ~2.58 GHz with a large coupling value reported in Ref. 15 could be excited in CM2 and show effects on the submacropulse beam centroid motion downstream of the CM. We had previously found evidence for its role in CC1 [4]. This mode would need to be near resonant with a beam harmonic to have much effect based on those studies. The frequency-integrated Schottky data for CM2:C1 US HOM showed a relatively strong signal, but no correlation with the HOMs. We suspect some monopole modes in this broad passband dominate the total signal and obscure the dipolar mode effect. The high-resolution frequency data in Fig. 16 do show modes whose amplitudes are sensitive to either the vertical (V) or horizontal (H) steering at 2.558 (V), 2.561 (H), 2.568 (V), and 2.572 (H) GHz which are indicated by the small black arrows in Fig. 16a.

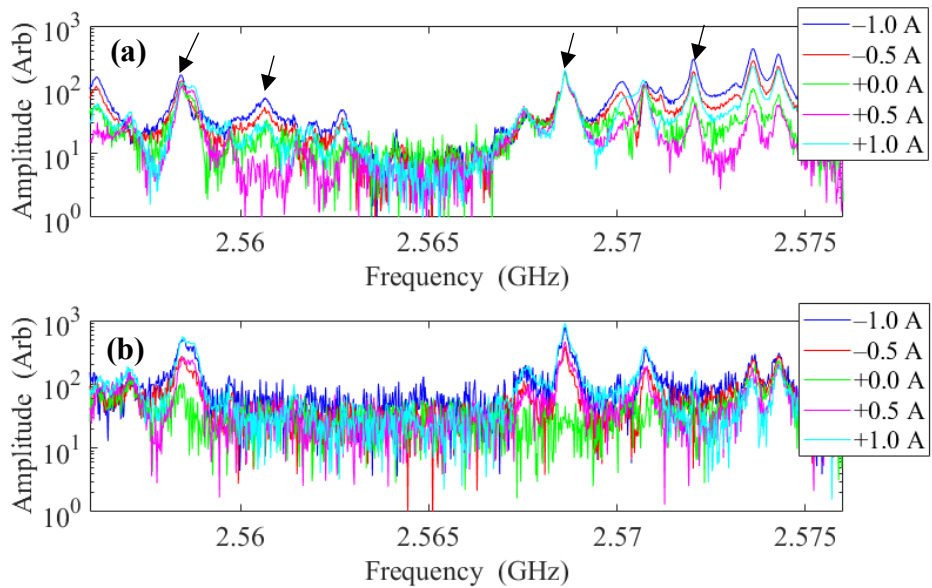

Figure 16: Horizontal (a) and vertical (b) steering results showing polarization-dependent HOM variations in several modes from 2.555 to 2.575 GHz in CM2:C1 as indicated by the arrows.

### E.3    3.25-GHz Passband Data (Integrated)

The objective for studies in this 3.25-GHz regime was to assess the quadrupole modes reported in a TESLA-type cavity [15]. The quadrupole modes are expected to be excited only by spatially asymmetric beam shapes inside the cavities. In Fig.17a we show that the C1 US HOM signal remains the same to a few mV (although non-zero) during the V125 scan while the C8 US HOM indicates some changes in excitation of the modes in this passband with V125 corrector steering by more than ±1 A or ± 2 mrad. In Fig. 17b we show the DS HOMs are excited for these two cavities. This suggests the beam shape was sufficiently asymmetric entering C1 US for the non-zero response, but there are focusing effects in C1 that change the shape before the C1 DS coupler. Both C8 US and C8 DS HOMs show a steering dependence presumably due to the asymetrical HOM quadrupole fields excited within the cryomodule. The signals for each corrector value were averaged over 300 shots with 250 pC/ b and 50 b.

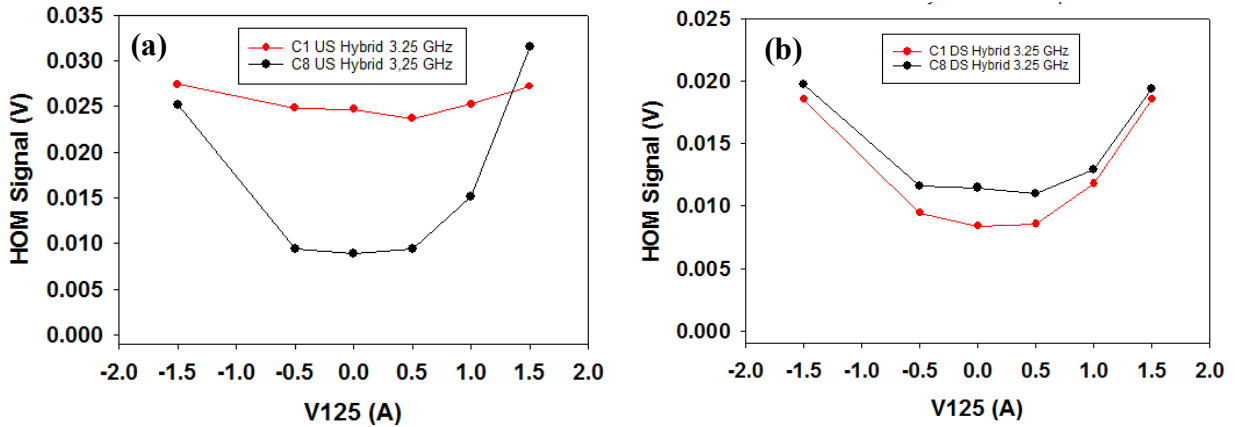

Figure 17: A comparison of the 3.25-GHz signals in the (a) CM2:C1 US HOM (red data) and the C8 US HOM (black data) and (b) C1 DS and C8 DS during the V125 scan. The data imply beam shape asymmetry entering C1, but a change in shape at the end of C1 and entering C8 for the larger off-axis steerings.

## IV. MACHINE LEARNING TRAINING

Machine Learning is undergoing a renaissance in a wide variety of applications due to larger computational resources, advanced theoretical models, and successful practical applications during the last years. Particle accelerators are part of this resurgence of ML due to developments of new system modelling techniques, virtual instrumentation/diagnostics, tuning and control schemes, surrogate models, among others [17]. In this section, we evaluate a neural network (NN) model for bunch-by-bunch centroid slewing prediction, and its application in a ML-based optimization and model construction for HOM signal level reduction and emittance preservation.

### A. Data source and structure

Our new experimental results in Section III showed a correlation between the electron-beam-induced cavity HOM signal levels and bunch-by-bunch centroid slewing and oscillation at the first 4 of 11 BPMS located downstream of the CM [18]. Thus, by reducing HOM signals and bunch-by-bunch centroid slewing downstream of the CM, one can mitigate emittance dilution. In principle, a NN control policy could be used for this purpose.

Waveform examples of US HOMs for all 8 cavities are shown in Fig. 4b. Although several features can be extracted from each of these waveforms (rise time, oscillation frequency, decay time), we decided to use the peak value as a representative number. Averaging the peak value over 300 shots, the relation between the V125 corrector current and HOM signal peaks at 400 pC/b and 50 bunches was shown in Fig. 8. The evolution of the relative beam centroid position as measured by B441PV (a BPM located ~1.2 m DS of the CM) over a 250 pC/b beam with 50 bunches is shown in Fig. 6 for multiple values of V125 corrector current and H125 at reference. A clear slew is present in the centroid position measurements which is proportional to the V125 corrector current and beam offset. With these results we can see how both HOM signal peaks and centroid slews are proportional to the corrector current (i.e., beam off-axis). In principle, we can train a NN to predict the centroid slew based on the HOM signal peaks.

### B. Neural Network Model

A NN was trained to predict the centroid motion's standard deviation as measured by multiple BPMs (B440PV/PH and B441PV/PH) over beams of 50 bunches, for several values of bunch charge and H/V125 corrector currents. The inputs to the NN are the US and DS HOM signal peaks as measured by the SLAC HOM detectors. The training data included measurements for beam charges of 200 and 100 pC/b, and H/V125 corrector currents from -1.0 A to +1.0 A from the reference current, with 0.5-A current steps. At each beam configuration, signals for 300 shots were captured. The result is a dataset with 6000 data points.

The NN architecture consisted of a normalization layer followed by 6 hidden layers (four layers of 100 nodes followed by two layers of 64 nodes). Each hidden layer used the hyperbolic tangent activation function. An 80-20 split was used for the training (4800 data points) and test datasets (1200 data points). From the training data set, 20% was used for validation. Early stop was implemented.

### C. Training Results

The performance of the model was evaluated in terms of the mean absolute error (MAE) and the mean absolute percentage error (MAPE). Computing resources of the SLAC Shared Scientific Data Facility (SDF) were used to perform the NN training [6]. The results are shown in Table II. The accuracy of the prediction of the standard deviation of the bunch-by-bunch centroid slew is about 8% for all BPMs shown. A representative histogram of the test data set MAPE for B441PV is shown in Fig. 18.

Table II: NN Performance Results

| BPM | Train MAE | Val MAE | Test MAE | Test MAPE |
|---|---|---|---|---|
| B440PV | 41.42 | 41.98 | 42.82 | 9.76 |
| B440PH | 29.82 | 30.46 | 30.54 | 8.2 |
| B441PV | 18.98 | 19.26 | 19.5 | 8.4 |
| B441PH | 20.89 | 21.43 | 21.62 | 8.44 |

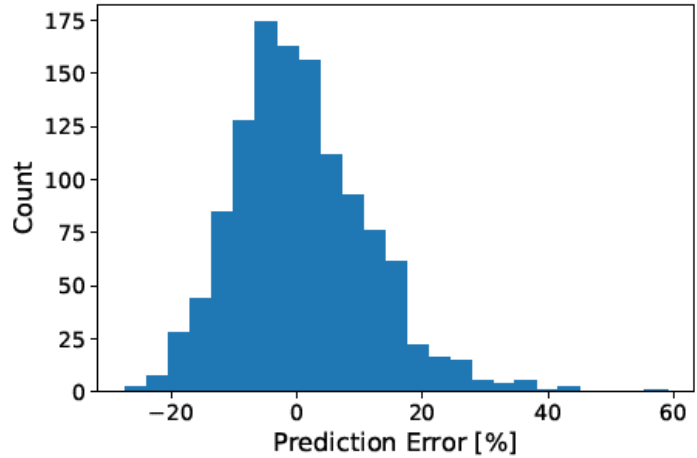

Figure 18: Histogram of prediction errors for B441PV.

The performance of the NN model for predictions of B441PV standard deviation over the test dataset is shown in Fig. 19. The x-axis represents the test dataset, which consists of 1200 data points (20% of the total dataset size). The predictions (gold) appear centered in the BPM real values (blue). The groups in Fig. 19 represent BPM measurements over the same beam and corrector configuration (i.e., fixed bunch charge and H/V125 corrector currents). These groups consist of ~60 data points, since we took 300 shots for each beam and corrector configuration. The NN model is capable of predicting the average bunch-by-bunch centroid slew's standard deviation for a given beam and corrector configuration. However, it is not accurate when predicting the exact value. This may be related to the noise on the BPM measurements and the low charge. Having the average bunch by bunch centroid slew's standard deviation might be enough when designing a controller based on these predictions.

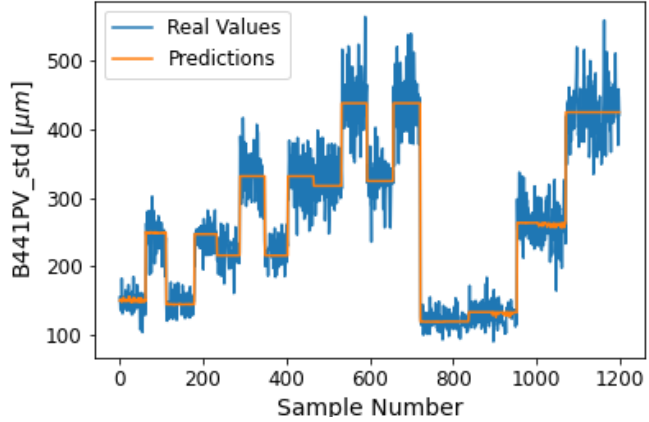

Figure 19: NN model predictions and real values for B441PV.

Data with the unprecedented correlation between beam steering, US and DS HOM signals, and BPM measurements showing bunch by bunch centroid slew after a Tesla-type CM at FAST have been used to train a NN model. Results show that the NN model is capable of predicting centroid slew's standard deviation with about 8% accuracy. These are encouraging results towards developing a ML-based controller for HOM reduction and emittance preservation for the LCLS-II project at SLAC. Our next steps include the development of the controller using an inverse model of the NN developed in this research, i.e., a NN that can predict HOM signals for a given beam offset. We also plan to explore adaptive learning, Gaussian Processes, and Gaussian Process based Bayesian optimization.

## V. SUMMARY

In summary, we have performed our first comprehensive study of the beam dynamics associated with HOMs in a TESLA-type cryomodule at the FAST/IOTA facility. We used both integrated HOM dipole mode responses online and high-resolution spectral measurements processed offline showing phase and frequency splitting of the polarization components of key modes. These were used to support HOM minimization and concomitant emittance-dilution reduction. This has been augmented by initial ML training with these data in support of this objective and for informing the commissioning of the LCLS-II injector and accelerator. These results also could have relevance to the EuXFEL injector where ~5-MeV beam enters the first CM. It is hoped that the data may be used to benchmark a simulation of such dynamics for a full cryomodule in the future.

**ACKNOWLEDGEMENTS**

The Fermilab authors acknowledge: the support of C. Drennan, A. Valishev, D. Broemmelsiek, G. Stancari, and M. Lindgren, all in the Accelerator Division at Fermilab; discussions with V. Yakovlev; and the CM2 network analyzer measurements by A. Lunin both of the Applied Physics and Superconducting Technology Division at Fermilab. The SLAC/NAL authors acknowledge the support of J. Schmerge (Superconducting Linac Division, SLAC). The Fermilab authors also acknowledge the support of Fermi Research Alliance, LLC under Contract No. DE-AC02-07CH11359 with the United States Department of Energy. The SLAC/NAL authors acknowledge support under contract DE-AC02-76SF00515 with the U.S. Department of Energy.